\documentclass[journal]{IEEEtran}

\ifCLASSINFOpdf
\else
   \usepackage[dvips]{graphicx}
\fi
\usepackage{url}

\usepackage{graphicx}
\usepackage{amsmath,amssymb,graphicx}
\usepackage[]{algorithm2e}
\usepackage{tabularx}
\usepackage{wasysym}
\usepackage{color,soul}
\usepackage{cite}
\usepackage{balance}
\usepackage{booktabs}
\usepackage{multirow}
\usepackage{siunitx}
\usepackage{hyperref}

\def\T{{\mathsf T}}

\def\RR{{\mathbb R}}
\usepackage{makecell}
\usepackage{caption}
\usepackage[font=scriptsize]{subfig}
\newcommand{\subparagraph}{}
\usepackage{titlesec}
\usepackage{textgreek}

\usepackage{soul}
\soulregister\cite7 %
\soulregister\citep7 %
\soulregister\citet7 %
\soulregister\ref7 %
\soulregister\pageref7 %
\newcommand{\ZQHL}[1]{{#1}} %

\begin{document}

\setlength{\abovedisplayskip}{3.5pt}
\setlength{\belowdisplayskip}{3.5pt}

\titlespacing*{\section}{0pt}{2.ex plus 0ex minus 0ex}{0.5ex plus 0ex minus 0ex}
\titlespacing*{\subsection}{0pt}{2.ex plus 0ex minus 0ex}{0.5ex plus 0ex minus 0ex}

\title{Improving Frame-Online Neural Speech Enhancement with Overlapped-Frame Prediction}

\author{Zhong-Qiu Wang and Shinji Watanabe
\thanks{Manuscript received on Mar. 14, 2022; revised Jun. 7, 2022.}
\thanks{Z.-Q. Wang and S. Watanabe are with Language Technologies Institute, Carnegie Mellon University, Pittsburgh, PA 15213, USA (e-mail: wang.zhongqiu41@gmail.com).}}

\markboth{}
{Shell \MakeLowercase{\textit{et al.}}: Bare Demo of IEEEtran.cls for IEEE Journals}

\maketitle

\begin{abstract}

Frame-online speech enhancement systems in the short-time Fourier transform (STFT) domain usually have an algorithmic latency equal to the window size due to the use of overlap-add in the inverse STFT (iSTFT).
This algorithmic latency allows the enhancement models to leverage future contextual information up to a length equal to the window size.
However, this information is only partially leveraged by current frame-online systems.
To fully exploit it, we propose an overlapped-frame prediction technique for deep learning based frame-online speech enhancement, where at each frame our deep neural network (DNN) predicts the current and \textit{several past frames} that are necessary for overlap-add, instead of only predicting the current frame.
In addition, we propose a loss function to account for the scale difference between predicted and oracle target signals.
Experiments on a noisy-reverberant speech enhancement task show the effectiveness of the proposed algorithms.

\end{abstract}


\IEEEpeerreviewmaketitle

\section{Introduction}

\IEEEPARstart{D}{eep} learning has elevated the performance of speech enhancement in the past decade \cite{WDLreview}.
Since the very first success of deep learning in offline enhancement \cite{WYXscaling}, there have been growing interests in using DNNs for low-latency enhancement, %
as many applications require online real-time processing.
For example, the recent deep noise suppression challenges \cite{K.A.Reddy2020} target at enhancement in a monaural teleconferencing setup, requiring a processing latency less than 40 ms on a specified processor.
Similar latency requirements exist in other related challenges \cite{Cutler2022, Rao2021}. %
The recent Clarity challenge \cite{ClarityWebpage} aims at multi-microphone enhancement in a hearing aid setup, requiring an algorithmic latency of at maximum 5 ms.

\begin{figure}
  \centering  
  \includegraphics[width=6cm]{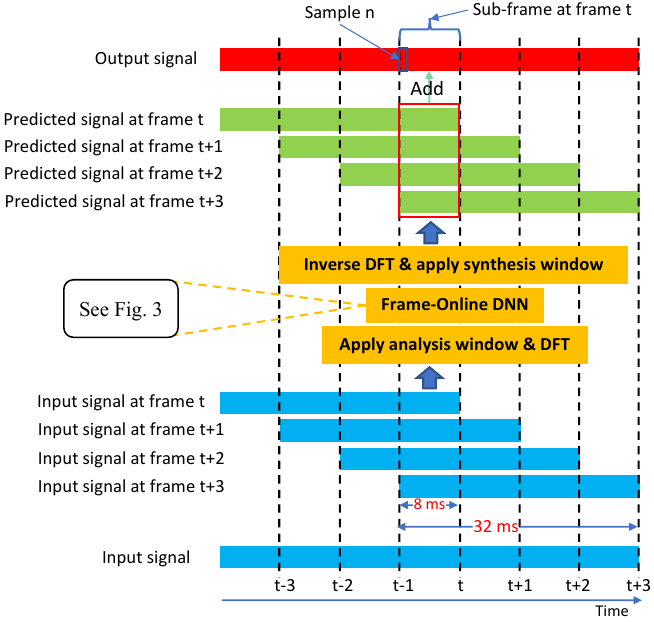}
  \vspace{-0.2cm} \\
  \caption{Single-frame prediction. Best viewed in color.}
  \label{single_frame_prediction}
  \vspace{-0.6cm}
\end{figure}

Numerous deep learning based approaches \cite{Chen2017b, Wichern2017, Luo2017a, Wilson2018, Wisdom2018, Higuchi2018, Luo2019, Yoshioka2019, Chakrabarty2019, Sonning2020, Liu2020, Tan2020, Hu2020, Pandey2020, Han2020, Xia2020, Defossez2020, Braun2021, Hao2021, Li2021, Tu2021, Yang2021, Zmolikova2021, Ren2021, LiChenda2022} have been proposed for frame-online speech enhancement by using frame-online (or causal) DNN modules such as uni-directional recurrent networks, causal normalization layers, causal convolutions and causal attention layers.
To the best of our knowledge, almost all the current DNN models adopt a \textit{single-frame prediction} strategy for online enhancement.
That is, the DNN models predict a frame of target speech at the current frame based on the current and past frames, and the prediction at the current frame is overlap-added with the predictions at nearby frames for signal re-synthesis.
This leads to an algorithmic latency equal to the window length.
However, this strategy cannot fully leverage the future context afforded by the algorithmic latency.
To explain this problem, we use an example STFT-based system shown in Fig.~\ref{single_frame_prediction} for illustration, where the window size is 32 ms and hop size 8 ms.
The frame-online DNN model operates in an online streaming fashion, processing one frame at a time when a new frame arrives and producing a predicted frame of signal for each input frame.
The predicted signals at nearby frames are then overlapped and added together (see the red-frame rectangle) to get the final output signal for the sub-frame at frame $t$ (marked in the top of the figure).
As we can see, to get the prediction at sample index $n$ (marked in the top), the system has to first fully observe the input signal at frame $t+3$ before the DNN can perform feed-forwarding.
The algorithmic latency is hence equal to the 32 ms window length.
In this system, to get the predicted signal at frame $t$, the model only takes in the input signals up to frame $t$.
The insight of this paper is that we can actually use input signals up to frame $t+3$ for the DNN to get the predicted signal at frame $t$ as well as at frame $t+1$, $t+2$ and $t+3$.
The resulting predicted signals would very likely be better, because the DNN can leverage up to three frames of future context, and at the same time the algorithmic latency remains the same as the window size.
This can be achieved by training our DNN model for \textit{overlapped-frame prediction}, where the current and past frames necessary for overlap-add are predicted at each frame at once.

Besides overlapped-frame prediction, we also propose a novel mechanism that equalizes the scales of predicted and oracle target signals before loss computation.

\section{Proposed Algorithms}\label{proposeddescription}

Let us denote the monaural mixture $y=s+v$ in the time domain as a summation of the target signal $s$ and non-target signal $v$.
In the STFT domain, we denote the mixture as $Y=S+V$, where $Y$, $S$ and $V$ are the STFT spectra of $y$, $s$ and $v$, respectively.
Our DNNs predict $S$ based on $Y$, if operating in the STFT domain; and predict $s$ based on $y$, if in the time domain.
This section describes the proposed overlapped-frame prediction, along with synthesis window design and DNN configurations, and its extensions to time-domain models.

\subsection{Overlapped-Frame Prediction}\label{ofp}

Let us denote the the STFT window and hop sizes by $W$ and $H$ samples, assuming $W$ is a multiple of $H$, and let $C=W/H$ be the number of overlapped frames in each window.
The proposed system is illustrated in Fig.~\ref{multi_frame_prediction}, which uses $32$ ms window and $8$ ms hop sizes as an example.
Our DNN operates in a frame-online fashion and, at each frame, it predicts $C$ frames consisting of the current and $C-1$ immediate past frames.
To get the sub-frame output at frame $t$ (marked in the top of the figure), we overlap the $C$ frames of predicted signals produced at frame $t+C-1$, and add together the sub-frames marked by the red-frame rectangle (denoted as ``partial sub-frame summation'').
Clearly, the algorithmic latency is still equal to the window size, but we can leverage input signals up to frame $t+C-1$ to better predict each of the $C$ frames that are overlap-added to get the sub-frame output at frame $t$.
This algorithm requires our frame-online DNN model to have $C$ outputs.
Alternatively, we can overlap-add the sub-frames marked by the black-frame rectangle (denoted as ``full sub-frame summation'').
This alternative could lead to better performance as it summates more sub-frame predictions.

We point out that our approach is different from studies \cite{Yoshioka2019, Sonning2020} that look ahead extra frames and add extra latency to the window size.
It is also different from studies \cite{Chen2016c} that predict a symmetric window of frames of time-frequency masks at each frame, where the motivation was about output ensembling.

\subsection{Synthesis Window Design}\label{synwin}

The proposed partial and full sub-frame summation methods require proper synthesis windows that can achieve perfect reconstruction when used with an analysis window $g \in \RR^W$.

\begin{figure}
  \centering  
  \includegraphics[width=8.5cm]{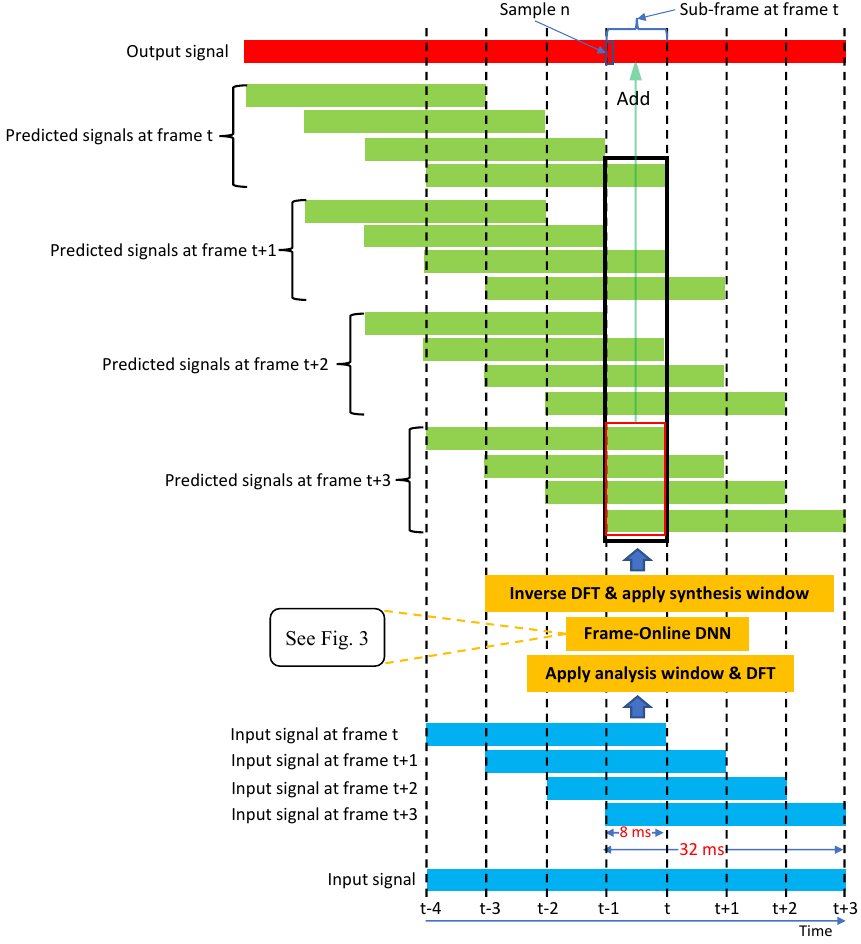}
  \vspace{-0.2cm} \\
  \caption{Overlapped-frame prediction. Best viewed in color.}
  \label{multi_frame_prediction}
  \vspace{-0.7cm}
\end{figure}

For the partial sub-frame summation, following \cite{Griffin1984a} we can use a regular synthesis window $l \in \RR^W$ defined as:
\begin{align}\label{reguarsynwin}
l[n] &= \frac{g[n]}{\sum_{e=0}^{W/H-1} {g[eH + (n \text{\,\,mod\,\,} H) ]}^2},
\end{align}
where $0 \leq n < W$.
\ZQHL{By following the overlap-add procedure, we can verify that perfect reconstruction is satisfied, because
\begin{align}\label{perfectreconstruction_partial}
\sum\nolimits_{e=0}^{W/H-1} {g[eH+h]\times l[eH+h]} = 1,
\end{align}
for any $0 \leq h < H$.}

For the full sub-frame summation, we design the synthesis window as follows:
\begin{align}\label{proposedsynwin}
l[n] &= \frac{g[n]}{\sum_{e=0}^{W/H-1} \Big((e+1) \times {g[eH + (n \text{\,\,mod\,\,} H)]}^2\Big)},
\end{align}
where the difference from Eq.~(\ref{reguarsynwin}) is that we have a weighting term for the $W/H$ sub-frames in the denominator.
\ZQHL{An intuitive way to understand this is} that among all the sub-frames marked by the black-frame rectangle in Fig.~\ref{multi_frame_prediction}, four are the last sub-frame in various predicted frames, three are the second-last sub-frame, and so on. 
We therefore need the weighting term in the denominator to have a pair of analysis and synthesis windows that can achieve perfect reconstruction, \ZQHL{which can be ensured, because, in the proposed overlap-add procedure, the condition below always holds for any $0 \leq h < H$:
\begin{align}\label{perfectreconstruction_full}
\sum\nolimits_{e=0}^{W/H-1} {g[eH+h]\times \Big( (e+1) \times l[eH+h] \Big)} = 1.
\end{align}
}

\subsection{DNN Configurations}\label{csm}

Our STFT-domain DNN model is trained for complex spectral mapping \cite{Williamson2016, Fu2017, Tan2020, Wang2020aCSMCHiME4, Wang2020CSMDereverbJournal}, where the real and imaginary (RI) components of the mixture $Y$ are concatenated as input features for a DNN to predict the RI components of target speech $S$.
The DNN architecture will be described in Section~\ref{dnnarchitecture}.

When trained for overlapped-frame prediction, the DNN predicts $K$ ($=C$) frames at each frame, essentially producing $C$ estimated target spectrograms; and when trained for single-frame prediction, the DNN produces one estimated target spectrogram (i.e. $K=1$).
Let us denote the DNN-estimated RI components by $\hat{R}^{(k)}$ and $\hat{I}^{(k)}$, where $k \in \{1,2,\dots,K\}$ indexes the $K$ spectrograms, and the enhanced speech by $\hat{S}^{(k)}=\hat{R}^{(k)}+j\hat{I}^{(k)}$, where $j$ is the imaginary unit.
We can define the loss function on the RI components and magnitudes of the estimated spectrograms:
\begin{align}\label{ri+mag}
\mathcal{L}_{\text{RI+Mag}} = \sum_{k}
\Big( \| \hat{R}^{(k)} - \text{Real}(S)\|_1 &+ \| \hat{I}^{(k)} - \text{Imag}(S)\|_1 \nonumber \\
&+ \Big\| |\hat{S}^{(k)}| - |S|\Big\|_1 \Big),
\end{align}
where $\text{Real}(\cdot)$ and $\text{Imag}(\cdot)$ extract RI components, $|\cdot|$ computes magnitude, and $\| \cdot\|_1$ calculates the $L_1$ norm.
Based on the described overlap-add mechanisms, an iSTFT is applied to re-synthesize the estimated time-domain signal $\hat{s}=\text{iSTFT}(\hat{S}^{(:)})$, where $\hat{S}^{(:)}$ includes all the $K$ estimated spectrograms.
Following \cite{Wang2021compensation}, we can train through the iSTFT and define the loss on the re-synthesized signal and its magnitude:
\begin{align}\label{wav+mag}
\mathcal{L}_{\text{Wav+Mag}} = & 
\| \hat{s} - s \|_1 + \Big\| |\text{STFT}(\hat{s})| - |\text{STFT}(s)| \Big\|_1,
\end{align}
where $\text{STFT}(\cdot)$ extracts a spectrogram from a signal\footnote{\ZQHL{We tried to weight the two terms in the loss function, but did not observe gains, likely because they have similar scales due to the Parseval's theorem.}}.
When using this loss function with overlapped-frame prediction, at each frame we essentially use signals up to frame $t+C-1$ to predict the sub-frame at frame $t$ (marked in the top of Fig.~\ref{multi_frame_prediction}).

When using mapping based approaches with the loss function in Eq.~(\ref{wav+mag}), the model needs to predict a signal that has the same gain as the target.
Although this may not be a problem for offline processing, as the model can observe the entire input mixture to produce a reasonable gain, for frame-online processing this could be difficult.
We propose to first compute a real-valued gain-equalization (GEQ) factor to balance the gain of the predicted signal with that of the target signal before loss computation:
\begin{align}\label{wav+mag+scale}
\mathcal{L}_{\text{Wav+Mag,geq}} = \| \hat{\alpha} \hat{s} &- s \|_1 + \nonumber \\
&\big\| |\text{STFT}(\hat{\alpha}\hat{s})| - |\text{STFT}(s)| \big\|_1,
\end{align}
where $\hat{\alpha}={{\text{argmin}}}_{\alpha}\,\| \alpha \hat{s} - s \|_2^2=(\hat{s}^{\T}s)/(\hat{s}^{\T}\hat{s})$.
\ZQHL{Note that GEQ can facilitate model training by not restricting the same gain, but $\hat{\alpha}$ will not be one at run time. This would be fine, as automatic gain adjustment, a common module in modern speech communication, can adjust the gain to a desired level.}

We always include in each loss function a magnitude loss, which can improve metrics that favor estimated signals with a good magnitude \cite{Wang2021compensation}.

\subsection{Extension to Time-Domain Models}\label{timedomainmodel}

\ZQHL{Besides STFT-domain models}, the proposed algorithms can also be used with time-domain models such as Conv-TasNet \cite{Luo2019}, by replacing the yellow blocks in Figs.~\ref{single_frame_prediction} and \ref{multi_frame_prediction} with time-domain models that use overlap-add for signal re-synthesis.

\section{Experimental Setup}\label{setup}

We validate our algorithms on a simulated monaural speech enhancement task.
This section describes the simulated dataset, DNN architectures, and miscellaneous configurations.

\subsection{Dataset}

The WSJCAM0 corpus \cite{Robinson1995} contains 7,861, 742, and 1,088 clean speech signals in its training, validation, and test sets, respectively.
Using the split of the clean signals in WSJCAM0, we simulate 39,245 ($\sim$77.7 h), 2,965 ($\sim$5.6 h) and 3,260 ($\sim$8.5 h) noisy-reverberant mixtures as our training, validation and test sets, respectively.
We sample the clips in the development set of FSD50k \cite{Fonseca2020} to simulate the noises for training and validation, and those in the evaluation set for testing.
For each simulated mixture, we sample up to 7 noise clips.
\ZQHL{We consider clips longer than 10 seconds as background noises and as foreground noises otherwise.
Each simulated mixed noise file has one background noise and the rest are foreground noises. 
The energy level between the dry background noise and each dry foreground noise is sampled from the range $[-3, 9]$ dB.}
The directions of each noise source and the target speaker to the microphone are independently sampled from the range $[0,2\pi)$.
We treat each sampled clip as a point source, convolve each source with its corresponding room impulse response, and summate the convolved signals to create the mixture.
After adding up all the spatialized noises, we scale the summated reverberant noise such that the signal-to-noise ratio between the target direct-path speech and the summated reverberant noise is equal to a value sampled from the range $[-8, 3]$ dB.
The distance from each source to the microphone is drawn from the range $[0.75, 2.5]$ m.
The reverberation time is drawn from the range $[0.2, 1.0]$ s.
The sampling rate is 16 kHz.

\begin{figure}
  \centering  
  \includegraphics[width=7.5cm]{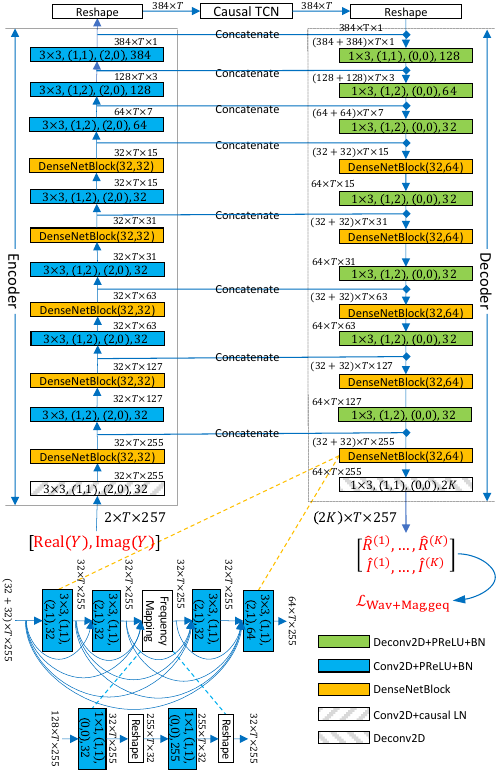}
  \vspace{-0.1cm} \\
  \caption{
  Frame-online TCN-DenseUNet architecture adapted from \cite{Wang2021LowDistortion}.
  During training, the tensor shape after each block 
  is denoted in the format: \textit{featureMaps}$\times$\textit{timeSteps}$\times$\textit{freqChannels}.
  Each one of Conv2D+PReLU+BN, Deconv2D+PReLU+BN, Conv2D, and Deconv2D blocks is shown in the format: \textit{kernelTime$\times$kernelFreq}, \textit{(strideTime, strideFreq)}, \textit{(padTime, padFreq)}, \textit{featureMaps}.
  \ZQHL{Each DenseNetBlock($g_1$,$g_2$) has five Conv2D+PReLU+BN blocks with growth rate $g_1$ for the first four layers and $g_2$ for the last one.}
  }
  \label{dnnfigure}
  \vspace{-0.6cm}
\end{figure}

\subsection{DNN Architectures}\label{dnnarchitecture}

Our STFT-domain frame-online DNN architecture is illustrated in Fig.~\ref{dnnfigure}.
It is a modified version of the offline TCN-DenseUNet architecture, which has shown strong performance in our previous studies \cite{Wang2020aCSMCHiME4, Wang2021FCPjournal, Wang2020css, Wang2021LowDistortion} on speech separation.
The major modifications include using causal layer normalization (LN) layers, batch normalization (BN) layers, and causal one- and two-dimensional convolution and deconvolution.
The network is a causal temporal convolution network (TCN) sandwiched by a U-Net, where the encoder performs down-sampling along frequency and the decoder performs up-sampling along frequency to recover the original frequency resolution.
DenseNet blocks, \ZQHL{which concatenate the outputs of the previous convolutions as the input to the current convolution}, are inserted at multiple frequency scales in the U-Net.
\ZQHL{The corresponding offline architecture is obtained by replacing causal LN and convolutions with non-causal ones.}
We choose this architecture as it shares many similarities with the architectures in recent studies \cite{Tan2020, Hu2020, Liu2020, Pandey2020}.
The RI components of different input (and output) signals are concatenated as feature maps in the network input (and output).
The number of feature maps is $2$ for the input tensor and is $2K$ for the output tensor.
Linear activation is used in the output layer to get the predicted RI components.
\ZQHL{Based on this architecture, overlapped-frame prediction uses only slightly more computation than single-frame prediction, since the increased computation, which only stems from the final 2D deconvolution and overlap-add, is negligible compared to the computation of the DNN.}

\ZQHL{We choose Conv-TasNet \cite{Luo2019} as the time-domain model for overlapped-frame prediction, considering its popularity.}
Its default hyper-parameters reported in \cite{Luo2019} are used.

\subsection{Miscellaneous Configurations}

\ZQHL{The direct-path signal of the target speaker is used as the target for model training and as the reference for metric computation.}
SI-SDR \ZQHL{in decibel (dB)} \cite{LeRoux2018a}, PESQ \cite{Rix2001}, and eSTOI \cite{Taal2011} are used as the evaluation metrics.
For TCN-DenseUNet, we experiment with the commonly used 32/8 and 20/10 ms window/hop sizes.
Giving a 16 kHz sampling rate, we use a 512-point discrete Fourier transform to obtain 257-dimensional STFT features at each frame for both setups.
This way, the same DNN architecture can be used.
The square-root Hann window is used as the analysis window.
For Conv-TasNet, we experiment with 4/2 ms window/hop sizes.
It should be noted that a longer window looks ahead more future samples and has a higher algorithmic latency, and would likely have more improvement when used with overlapped-frame prediction.

\begin{table}[t]
\scriptsize
\centering
 \sisetup{table-format=2.2,round-mode=places,round-precision=2,table-number-alignment = center,detect-weight=true,detect-inline-weight=math}
\caption{\textsc{\scriptsize Results of using 32/8 ms Window/Hop Sizes for TCN-DenseUNet.}}
\vspace{-0.15cm}
\label{results32and8}
\setlength{\tabcolsep}{2pt}
\begin{tabular}{
c
l
ccc
S[table-format=2.1,round-precision=1]
S[table-format=1.2]
S[table-format=1.3,round-precision=3]
}
\toprule
&           & Frame- & Loss & Sub-frame & {SI-} & & \\

ID & Systems & online? & function & summation & {SDR} & {PESQ} & {eSTOI} \\

\midrule

0 & Unprocessed & - & - & - & {$-$6.2} & 1.444 & 0.41129 \\

\midrule

1a & Single-frame pred. & yes & RI+Mag & - & 3.3623 & 2.111 & 0.70803 \\
1b & Overlapped-frame pred. & yes & RI+Mag & Partial & 3.63533 & 2.181714 & 0.725149 \\
1c & Overlapped-frame pred. & yes & RI+Mag & Full & 3.5685 & 2.157097 & 0.72230 \\

\midrule

2a & Single-frame pred. & yes & Wav+Mag & - & 3.464 & 2.1693 & 0.719478 \\
2b & Overlapped-frame pred. & yes & Wav+Mag & Partial & 4.0031 & 2.2270 & 0.73284 \\
2c & Overlapped-frame pred. & yes & Wav+Mag & Full & 3.9672 & 2.25643 & 0.7382654 \\
2d & Overlapped-frame pred. & yes & Wav+Mag,geq & Full & 4.17728 & 2.27988 & 0.74291 \\

\midrule

3a & Single-frame pred. & no & RI+Mag & - & 4.8089 & 2.44499 & 0.775965 \\
3b & Single-frame pred. & no & Wav+Mag & - & 4.62661 & 2.4573953 & 0.77563106 \\
3c & Single-frame pred. & no & Wav+Mag,geq & - & 4.8077828 & 2.468124 & 0.7776754 \\

\bottomrule
\end{tabular}
\vspace{-0.4cm}
\end{table}

\section{Evaluation Results}\label{results}

Table~\ref{results32and8}, \ref{results20and10} and \ref{resultstime4and2} respectively report the results of using 32/8, 20/10 and 4/2 ms window/hop sizes. 
In each table, we provide the offline results obtained by using the offline versions of the DNN architectures.
These offline results can be viewed as the performance upper bound of the frame-online models.

\begin{table}[t]
\scriptsize
\centering
 \sisetup{table-format=2.2,round-mode=places,round-precision=2,table-number-alignment = center,detect-weight=true,detect-inline-weight=math}
\caption{\textsc{\scriptsize Results of using 20/10 ms Window/Hop Sizes for TCN-DenseUNet.}}
\vspace{-0.15cm}
\label{results20and10}
\setlength{\tabcolsep}{2pt}
\begin{tabular}{
c
l
ccc
S[table-format=2.1,round-precision=1]
S[table-format=1.2]
S[table-format=1.3,round-precision=3]
}
\toprule
&           & Frame- & Loss & Sub-frame & {SI-} & & \\

ID & Systems & online? & function & summation & {SDR} & {PESQ} & {eSTOI} \\

\midrule

1a & Single-frame pred. & yes & RI+Mag & - & 2.817165 & 2.00948 & 0.6915467 \\
1b & Overlapped-frame pred. & yes & RI+Mag & Partial & 3.040456 & 2.04749 & 0.698827 \\
1c & Overlapped-frame pred. & yes & RI+Mag & Full & 2.97353 & 2.0501 & 0.69835 \\

\midrule

2a & Single-frame pred. & yes & Wav+Mag & - & 2.30509 & 2.061385 & 0.701681 \\
2b & Overlapped-frame pred. & yes & Wav+Mag & Partial & 2.4047 & 2.062925 & 0.70361 \\
2c & Overlapped-frame pred. & yes & Wav+Mag & Full & 3.00844 & 2.157637 & 0.721326 \\
2d & Overlapped-frame pred. & yes & Wav+Mag,geq & Full & 3.43433 & 2.159852 & 0.724427 \\

\midrule

3a & Single-frame pred. & no & RI+Mag & - & 4.533 & 2.46115 & 0.770312 \\
3b & Single-frame pred. & no & Wav+Mag & - & 3.9643 & 2.46524763 & 0.77833 \\
3c & Single-frame pred. & no & Wav+Mag,geq & - & 4.506760 & 2.485129 & 0.783984 \\

\bottomrule
\end{tabular}
\vspace{-0.25cm}
\end{table}

Let us first look at Table~\ref{results32and8}.
Comparing 1a and 1b (or 2a and 2b), we observe that overlapped-frame prediction with partial sub-frame summation produces better performance than single-frame prediction.
Using full sub-frame summation, 1c obtains worse performance than 1b, while 2c obtains \ZQHL{similar or slightly better} performance than 2b.
This is likely because when we use the RI+Mag loss and do not train through iSTFT, the sub-frames summated by using full sub-frame summation contain sub-frame predictions produced at earlier frames (see the black-frame rectangle in Fig.~\ref{multi_frame_prediction}), which are not as good as the ones produced at the current frame.
When we train through iSTFT and use the Wav+Mag loss, the model may figure out how to best predict each sub-frame and best summate all the sub-frames (in the black-frame rectangle) to optimize the loss.
Overall, our proposed systems in 2c show noticeable improvement over 1a and 2a, by better leveraging the 32 ms future context information.

In Table~\ref{results20and10}, similar trend as in Table~\ref{results32and8} is observed.
The relative gains of overlapped-frame prediction over single-frame prediction are smaller than those in Table~\ref{results32and8}.
This is as expected as there is less future context (i.e., 20 ms) to exploit.

In Table~\ref{resultstime4and2}, the gains are very marginal as the future context (i.e., 4 ms) is even less. \ZQHL{This Conv-TasNet experiment suggests that the window size needs to be reasonably large for overlapped-frame prediction to work well.} %

Although the gains brought by overlapped-frame prediction depend on the allowed future context and are small when the allowed context is limited, we always observe consistent and steady improvements in the considered setups.

\begin{table}[t]
\scriptsize
\centering
 \sisetup{table-format=2.2,round-mode=places,round-precision=2,table-number-alignment = center,detect-weight=true,detect-inline-weight=math}
\caption{\textsc{\scriptsize Results of using 4/2 ms Window/Hop Sizes for Conv-TasNet.}}
\vspace{-0.15cm}
\label{resultstime4and2}
\setlength{\tabcolsep}{2pt}
\begin{tabular}{
c
l
ccc
S[table-format=2.1,round-precision=1]
S[table-format=1.2]
S[table-format=1.3,round-precision=3]
}
\toprule
&           & Frame- & Loss & Sub-frame & {SI-} & & \\

ID & Systems & online? & function & summation & {SDR} & {PESQ} & {eSTOI} \\

\midrule

2a & Single-frame pred. & yes & Wav+Mag & - & 2.1758236 & 1.790139 & 0.6582973 \\
2b & Overlapped-frame pred. & yes & Wav+Mag & Partial & 2.109716 & 1.774767 & 0.6523726 \\
2c & Overlapped-frame pred. & yes & Wav+Mag & Full & 2.265678 & 1.809428 & 0.66109 \\
2d & Overlapped-frame pred. & yes & Wav+Mag,geq & Full & 2.3217534 & 1.8096339 & 0.66398 \\

\midrule

3b & Single-frame pred. & no & Wav+Mag & - & 3.8323413 & 2.275824 & 0.753824 \\
3c & Single-frame pred. & no & Wav+Mag,geq & - & 3.86827392 & 2.25686 & 0.75385 \\

\bottomrule
\end{tabular}
\vspace{-0.5cm}
\end{table}

In all the tables, the proposed gain equalization leads to slightly better results in the online cases (see 2c vs. 2d).

\section{Conclusion}\label{conclusion}

We have proposed a novel overlapped-frame prediction technique%
for frame-online speech enhancement.
It better leverages future context, without incurring extra algorithmic latency.
The increased amount of computation only stems from the output layer and is negligible compared to that of the DNN backbone.
The proposed technique can be easily modified for, or directly adopted by, numerous frame-online systems.
It would likely yield better performance due to the better exploitation of the future context afforded by the algorithmic latency, no matter whether they are DNN- or non-DNN-based, whether they operate in the T-F domain or in the time domain, or whether they deal with speech enhancement or other related separation tasks, as long as they use overlap-add for signal re-synthesis.

\bibliographystyle{IEEEtran}
\bibliography{references.bib}

\end{document}